\documentclass[preprint,showpacs]{revtex4} 

\usepackage{graphicx}

\usepackage{dcolumn}
\usepackage{bm}

\begin{document}

\title{Studies of collisional dephasing of two-photon excited atomic sodium}

\author{Leonid Rybak, Lev Chuntonov, Andrey Gandman, Naser Shakour, and Zohar Amitay}
\email{amitayz@tx.technion.ac.il} %
\affiliation{Schulich Faculty of Chemistry, Technion - Israel
Institute of Technology, Haifa 32000, Israel}

\begin{abstract}

Coherence relaxation in Na vapor was studied utilizing a
"pump-probe" method with femtosecond pulses. The coherence was
created by the pump pulse, which excited the Na atoms to the 4S
state. The atoms were further excited to the 7P state by the probe
pulse. As a result a coherent UV radiation was emitted by the atoms.
This emission was used to study the relaxation of both the 4S-3S and
7P-3S coherent superpositions. The relaxation is mainly due to
collisions with the Ar atoms serving as a buffer gas. Coherence
relaxation times and collisional cross-sections for the Na-Ar
collision pairs were extracted experimentally.

\end{abstract}


\maketitle


The development of femtosecond pulses has led to a breakthrough in
experimental methods and technologies. These new techniques make it
possible to obtain detailed information on the dynamics of various
processes, which occur on a pico- and sub-picosecond timescale.
Ultrashort pulse duration allows for real-time monitoring of such
processes. The most widely used experimental method here is the
"pump-probe" technique \cite{1,2,3,4,5,6}. The pump pulse, which is
used to excite the system under study, is followed by the probe
pulse, which is applied after the time delay $\tau$ and detects the
changes occurring in the system during this time interval. The time
delay $\tau$ is varied by varying the optical path length of the
probe pulse. The systems response, which depends on this time delay,
is recorded via variuos methods including fluorescence, absorption,
rotation of polarization plane of the probe pulse, as well as
nonlinear methods using four-wave mixing, etc.

The study presented in this paper investigates coherence relaxation
in atomic vapors. The atomic coherence, which is excited by the pump
pulse, relaxes exponentially with the characteristic decay time
$T_{2}$\cite{7,8,9}. This dynamics occur on a picosecond timescale,
therefore requiring ultrashort (femtosecond) pulses to study it. In
this study femtosecond pulses are used to excite Na atoms to a
coherent superposition of the atomic field-free eigenstates. The
decay of this superposition is studied then utilizing the
resonance-mediated generation of coherent UV radiation.

\begin{figure} 
\includegraphics[width=17cm]{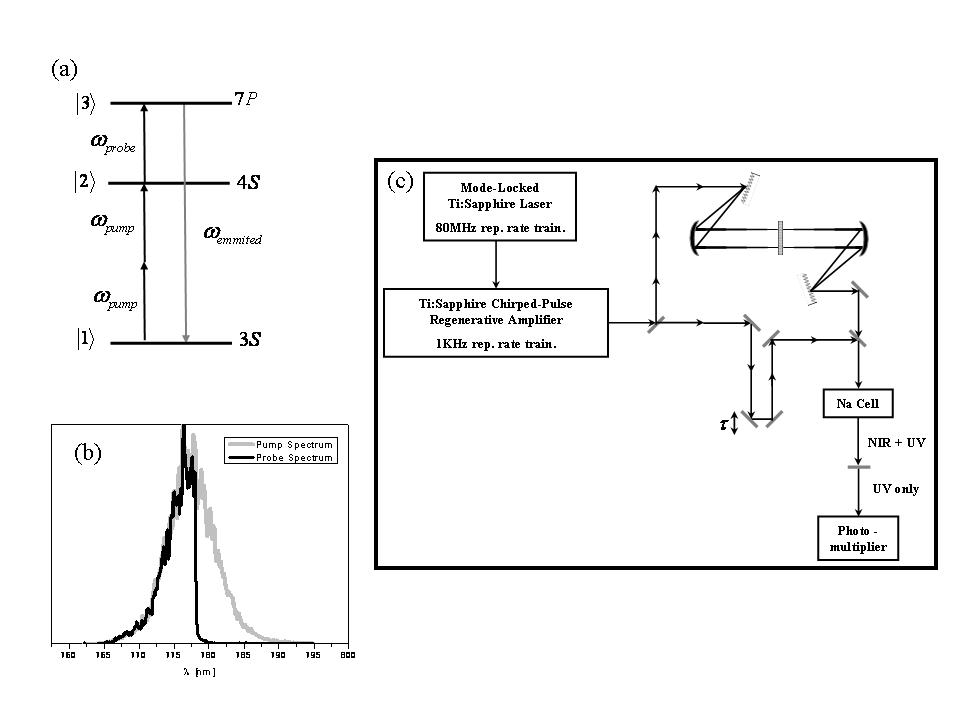}     
\vspace*{-0.7cm} \caption{(a) The model system. The 4S state is
excited by the pump pulse via a two-photon transition, followed by a
one photon excitation of the 7P state. (b) The spectra of the pump
(black) and the probe (gray) pulses. The pump pulse spectrum was
blocked at the low frequency side (see text). (c) Experimental
scheme.} \label{fig_1}
\end{figure}

Consider a three-level atom, with levels $\left|1\right\rangle$,
$\left|2\right\rangle$ and $\left|3\right\rangle$ as depicted in
Fig. \ref{fig_1}. States $\left|1\right\rangle$ and
$\left|2\right\rangle$ are of the same symmetry, while state
$\left|3\right\rangle$ is of a different symmetry. Suppose the atom
is irradiated by a pump-probe sequence of two collinear femtosecond
pulses. The pulses are separate in time with a time delay $\tau$. In
what follows both pulses are assumed to be much shorter then all the
relaxation times of the atom, and therefore will be treated as
$\delta$-function. The pump pulse, which is taken to be centered at
$t=0$, excites a coherent superposition of the states
$\left|1\right\rangle$ and $\left|2\right\rangle$, via a two-photon
transition. The probe pulse excites the state
$\left|3\right\rangle$, via a one-photon transition, thus creating a
coherent superposition of the states $\left|1\right\rangle$ and
$\left|3\right\rangle$. As a result an atomic dipole moment is
formed between the $\left|1\right\rangle$ and $\left|3\right\rangle$
states. This dipole moment oscillates at the
$\left|3\right\rangle$-$\left|1\right\rangle$ transition frequency
\cite{10}.

Now consider a collection of atoms described above. The atomic
dipoles will combine to form a matter polarization wave, which will
cause an emission of light at the
$\left|3\right\rangle$-$\left|1\right\rangle$ transition frequency,
$\omega_{3,1}$. This polarization is given by the expectation value
of the dipole moment operator, which can be calculated is follows:
\begin{equation}\label{eq_1}
    P(t,\tau)=\langle\mu\rangle=tr(\boldsymbol{\rho\mu})
\end{equation}
where $\boldsymbol{\rho}$ is the density matrix of the system and
$\boldsymbol{\mu}$ is the matrix representing the dipole moment
operator. When considering the selection rules of the atom (see
above), it turns out that only one density matrix element,
$\rho_{3,1}$, needed to calculate the polarization above, thus:
\begin{equation}\label{eq_2}
    P(t,\tau)=\rho_{3,1}(t,\tau)\mu_{3,1}+c.c.
\end{equation}
This matrix element is easily calculated, provided the pulses are
approximated by $\delta$-functions. It is given by:

\begin{equation}\label{eq_3}
\rho_{3,1}(t,\tau)\propto\begin{array}{cc}
                             0 & t\leq\tau \\
                             \exp[(i\omega_{2,1}-\gamma_{2,1})\tau]\cdot\exp[(i\omega_{3,1}-\gamma_{3,1})(t-\tau)] & t>\tau \\
                           \end{array}
\end{equation}
where $\omega_{2(3),1}$ is the
$\left|2(3)\right\rangle$-$\left|1\right\rangle$ transition
frequency and $\gamma_{2(3),1}=1/T^{2(3),1}_{2}$, with
$T^{2(3),1}_{2}$ being the relaxation time of the
$\left|2(3)\right\rangle$-$\left|1\right\rangle$ coherence. The
first exponential term in Eq. \ref{eq_3} describes the evolution of
the $\left|2\right\rangle$-$\left|1\right\rangle$ coherent
superposition created by the pump pulse at $t=0$, while the second
exponential term describes the evolution of the
$\left|3\right\rangle$-$\left|1\right\rangle$ coherent superposition
created by the probe pulse at $t=\tau$ ($\tau>0$). In atomic vapors
the relaxations times $T^{2(3),1}_{2}$ are determined by collisional
dephasing processes governed by the collisional cross section given
by:

\begin{equation}\label{eq_4}
    \sigma=\frac{1}{\widetilde{N}v_{r}T_{2}}
\end{equation}
where $v_{r}=(8kT/\pi\mu)^{1/2}$ is the mean relative velocity of
the collision pair, $\widetilde{N}=P/kT$ is the number density of
the colliding atoms, with $\mu$ being the reduced mass of the
collision pair, $T$ being the absolute temperature, $P$ being the
pressure and $k$ being the Boltzmann constant.

When dealing with atomic vapors, one must also consider the "Doppler
effect" when calculating the field generated by the polarization in
Eq. \ref{eq_2} \cite{11}. Since according to the "Doppler effect"
each atom has resonance frequencies, which depend on its velocity,
$v$, i.e,
$\omega_{n,m}(v)=\omega_{n,m}^{0}+\frac{v}{c}\omega_{n,m}^{0}$, the
vapor can be said to consist of various collections of atoms, while
each such collection is composed of atoms having the same velocity.
As a result each such collection of atoms creates its own
polarization given by modification of Eqs. \ref{eq_2}-\ref{eq_3}.
\begin{equation}\label{eq_5}
    P(t,\tau,v)\propto\exp[(i\omega_{2,1}(v)-\gamma_{2,1})\tau]\cdot\exp[(i\omega_{3,1}(v)-\gamma_{3,1})(t-\tau)]
\end{equation}
The total electric field of the emitted light will be given by a
coherent sum over various polarizations, each of which is created by
atoms of different velocity:
\begin{equation}\label{eq_6}
    E(t,\tau)\propto\int^\infty_\infty W(v)P(t,\tau,v)dv
\end{equation}
where $W(v)=\frac{1}{u\sqrt{\pi}}e^{-(\frac{v}{u})^{2}}$ is the
Boltzmann velocity distribution, with $u=\sqrt{\frac{2kT}{m}}$,
where $m$ is the mass of the atom.

As will be shown below the collisional cross sections (see Eq.
\ref{eq_4}) can be extracted experimentally by measuring the total
energy of the field in Eq. \ref{eq_6} as a function of the
pump-probe delay $\tau$, i.e, $S(\tau)=\int_{\tau}^{\infty} \mid
E(t,\tau)\mid^2dt$.

The above theoretical scheme was implemented with Na atoms, with
argon serving as a buffer gas. The relevant energy levels of the Na
atom are shown in Fig. \ref{fig_1} with the 3S state serving as the
$\left|1\right\rangle$ state, the 4S state serving as the
$\left|2\right\rangle$ state and the 7P state serving as the
$\left|3\right\rangle$ state. The two-photon coupling between the 3S
and the 4S states is provided by the various P-states that are far
from resonance and are not shown. The corresponding transition
frequencies are $\omega_{4S,3S}=25740\ cm^{-1}$ (two $777\ nm$
photons) and $\omega_{7P,4S}=12801\ cm^{-1}$ (one $781.2\ nm$
photon). Experimentally atomic Na vapor is held in a heated cell at
a temperature of $400\ ^{o}C$ filled with Ar buffer gas.

The spectrum of the NIR femtosecond pulses used in the experiment is
shown in Fig. \ref{fig_1}. It is centered around $777.4\ nm$ with a
spectral bandwidth of $7.6\ nm$ (FWHM) (corresponding to $117 \ fs$
pulse duration). The initial pulse was divided into two replicas of
itself constituting the pump and the probe pulses. The pump pulse
was passed trough a 4f pulse shaping setup, where it had its
spectrum blocked at the low-frequency side (see Fig. \ref{fig_1}) to
eliminate resonant access to the 7P state\cite{10} ,thus the pump
populates the 4S state only, in correspondence with theoretical
scheme above. The probe pulse is sent trough a delay arm, where the
pump-probe delay $\tau$ is set. Then both pulses are recombined and
are focused in the Na cell. The UV emission, induced by the pulses,
is separated from the NIR excitation by a proper optical filter and
the total energy of this emission is then measured as a function of
$\tau$, thus corresponding to $S(\tau)$ described above (see
Fig\ref{fig_1} for the experimental setup).

As can be seen from Eqs.\ref{eq_5}-\ref{eq_6}, the rate of decay of
$S(\tau)$ with $\tau$ is governed by the coherence relaxation times
$T_{2}^{4S}$ and $T_{2}^{7P}$. The knowledge of these relaxation
times at different pressures should provide the collisional
cross-sections $\sigma_{4S}$ and $\sigma_{7P}$, between Ar and Na
atoms in 3S-4S and 3S-7P superpositions, respectively, according to
Eq.\ref{eq_4}.

\begin{figure} 
\includegraphics[width=17cm]{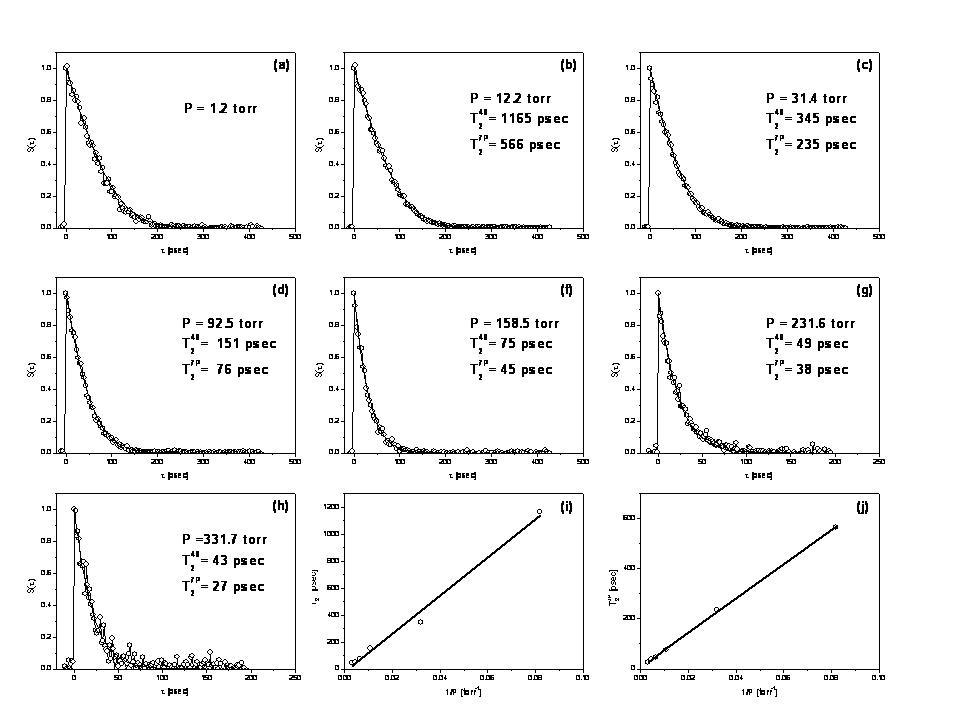}     
\vspace*{-0.7cm} \caption{(a)-(h) Experimental (circles) and
numerical (solid) results for $S(\tau)$ at different Ar pressures.
(i)-(j) Plots of $T_{2}^{4S(7P)}$ vs 1/P for determination of
$\sigma_{4S(7P)}$. The collisional cross-section $\sigma$ is
determined from the slope of the plots (see text).} \label{fig_2}
\end{figure}

Experimentally $S(\tau)$ was measured at different pressure of Ar.
First, $S(\tau)$ was measured at low Ar pressure (1.2 torr) to
verify the effect of Doppler broadening (see Eqs.
\ref{eq_5}-\ref{eq_6}). The experimental results are presented in
Fig. \ref{fig_2} (a) (circles) along with the numerical results
(solid line) calculated based on Eqs. \ref{eq_5}-\ref{eq_6}, where
the coherence decay constants $\gamma_{4S(7P),3S}$ were set to zero
and the temperature set to 400 $^{o}C$. The excellent agreement
between the measured and calculated results confirms the fact that
the main dephasing mechanism at low peasures is the inhomogenious
Doppler broadening. At the temperature of the experiment ($T=400
^{o}C$) it was calculated to be $0.1 cm^{-1}$ and $0.15 cm^{-1}$ for
the 4S-3S and for the 7P-3S transitions respectively.

The coherence relaxation times $T_{2}^{4S}$ and $T_{2}^{7P}$ were
determined experimentally by measuring $S_(\tau)$ for different Ar
pressures. The experimental results are presented in Fig.\ref{fig_2}
(b)-(h) (circles). By performing a least-squares fit between
numerical calculation, and the entire experimental data, the
relaxation times and the collisional cross-section were obtained.
The numerical results are shown in Fig.\ref{fig_2} (b)-(h) (solid
lines). The collisional cross-sections were extracted from the
slopes of the plots of $T_{2}^{4S(7P)}$ vs $1/P$ (acc. to Eq.
\ref{eq_4}). These plots are shown in Fig. \ref{fig_2} (i)-(j), and
the collisional cross-sections were found to be $\sigma_{4S} = 488
\pm 23.6 {\AA}^{2}$ and $\sigma_{7P}=1004\pm13.3{\AA}^{2}$. These
values agree qualitatively with previously measured
results\cite{7,12} for collisional cross-section between Ar and Na
atoms, excited to different levels.


In summary, we have used the nonlinear generation of UV emission to
probe the coherences in a Na atom. The calculation, based on the
theoretical model developed here, agree well with the experimental
results and numerical values for the coherence relaxation times were
obtained. Based on those, the collisional cross-section for for the
collisions of Ar and Na atoms in a 4S-3S and 7P-3S coherent
superpositions, were established.


\newpage

\newpage

\end{document}